\documentclass[aps,prb,twocolumn,floatfix, 10pt, superscriptaddress,eqsecnum,showpacs,showkeys,amsmath,amssymb]{revtex4}
\usepackage{amsfonts}
\usepackage{amssymb,amsmath}
\usepackage{graphicx}
\usepackage{dcolumn}
\usepackage{bm}
\usepackage{color}
\let\a=\alpha \let\be=\beta \let\g=\gamma \let\de=\delta
  \let\h=\eta 
\let\eps=\epsilon
   \let\m=\mu
   \let\r=\rho 
\let\om=\omega 
\let\ph=\varphi   
  
  \let\D=\Delta

\let\qd=\quad \let\qqd=\qquad 
\def\epp{\, .}
\def\epc{\, ,}

\def\tst#1{{\textstyle #1}}

\def\2{\frac{1}{2}} \def\4{\frac{1}{4}}
\def\6{\partial}
\def\+{\dagger}
\def\<{\langle} \def\>{\rangle}

\def\rd{{\rm d}}
\def\re{{\rm e}}

\DeclareMathOperator{\sh}{sh}
\renewcommand{\sinh}{\sh}
\DeclareMathOperator{\ch}{ch}
\renewcommand{\cosh}{\ch}

\DeclareMathOperator{\tr}{tr}

\DeclareMathOperator{\sign}{sign}

\def\Ev{\mathbf{E}}
\def\Hv{\mathbf{H}}
\def\Mv{\mathbf{M}}
\def\Pv{\mathbf{P}}

\def\fb{\mathfrak{b}}
\def\fbq{\overline{\mathfrak{b}}}

\newcommand{\mfb}{\mathfrak{b}}

\newcommand{\mfbq}{\overline{\mathfrak{b}}}

\newcommand{\mcH}{\mathcal{H}}

\newcommand{\mcU}{\mathcal{U}}

\newcommand{\nut}{\tilde{\nu}}

\renewcommand{\tilde}{\widetilde}

\newcommand{\Rarrow}[2][5mm]{\xrightarrow{\makebox[#1][c]{#2}}}

\begin{document}

\title{
Exact description of the  magnetoelectric effect in the
spin-1/2 XXZ-chain with Dzyaloshinskii-Moriya interaction}

\author{Michael Brockmann}
\affiliation{Institute of Physics,
             University of Amsterdam,
             Science Park 904, 1098 XH Amsterdam, Netherlands}
\author{Andreas Kl\"{u}mper}
\affiliation{Theoretische Physik,
             Bergische Universit\"{a}t Wuppertal,
             Gau{\ss}str.~20, D-42119, Wuppertal, Germany}
\author{Vadim Ohanyan}
\affiliation{Department of Theoretical Physics,
             Yerevan State University,
             Alex Manoogian 1, 0025 Yerevan, Armenia}
\affiliation{ICTP, Strada Costiera, 11 I-34151 Trieste, Italy}
\date{\today}

\begin{abstract}
We consider a simple integrable model of a spin chain exhibiting the
Magnetoelectric Effect (MEE). Starting from the periodic $S\!=\!1/2$
XXZ-chain with Dzyaloshinskii-Moriya terms, which we consider as a
local electric polarization in the spirit of the
Katsura-Nagaosa-Baladsky (KNB) mechanism, we perform the mapping
onto the conventional XXZ-chain with twisted boundary conditions.
Using the techniques of Quantum Transfer Matrix (QTM) and Non-Linear
Integral Equations (NLIE) we obtain the magnetization, electric
polarization and magnetoelectric tensor as functions of magnetic and
electric field for arbitrary temperatures. We investigate these
dependencies as well as the thermal behavior of the above mentioned
physical quantities, especially in the low-temperature regime. We
found several regimes of polarization. Adjusting the magnetic field
one can switch the system from one regime to another. The features
of the critical properties connected with the MEE are also
illustrated.
\end{abstract}



\maketitle

\section{Introduction}\label{sec1}
The strong interest in multiferroic materials (materials exhibiting
simultaneously several primary ferroic order parameters), which
occurred recently,\cite{mult1,mult2,mult3} also triggered intensive
investigations in the field of magnetoelectric effects (MEE) and
magnetoelectric materials.\cite{fiebig} In general, MEE denominates
the mutual influence of the electric and magnetic properties in
matter. In its most prominent form, MEE can be defined as the
magnetic field dependence of (ferro)electric polarization and the
electric field dependence of magnetization. In linear approximation,
one can introduce the linear magnetoelectric tensor (isothermic),
which quantifies the induction of (ferro)electric polarization by a
magnetic field and of magnetization by an electric field:
\begin{equation}\label{alpha}
  \alpha_{ij}=\left(\frac{\partial P_i}{\partial H_j}\right)_{T,\,\Ev}
  = \left(\frac{\partial M_j}{\partial E_i}\right)_{T,\,\Hv}\epc
\end{equation}
where $T$ is the temperature, $i,j=(x,y,z)$, $\Pv$ ($\Mv$) the
(ferro)electric polarization (magnetization) and $\Ev$ ($\Hv$) the
electric (magnetic) field, respectively.  The efficient control of
magnetic properties of solids by means of the electric field has a
lot of potential applications, for instance, in
spintronics.\cite{jia09} At the moment, the number of known
materials exhibiting the MEE exceeds one hundred.\cite{mult2} One of
the microscopic mechanisms of the MEE was theoretically described by
Katsura, Nagaosa, and Baladsky in 2005 (KNB-mechanism)\cite{KNB} and
connects the appearance of the local (ferro)electric polarization
with the non-collinear ordering of the neighboring spins,
i.\,e.~with the spin current:\cite{KNB,ser_dag,jia}
\begin{equation}\label{loc_pol}
  \mathbf{P}_i \sim \mathbf{e}_{ij}\times(\mathbf{S}_i\times\mathbf{S}_j)\epc
\end{equation}
where $\mathbf{e}_{ij}$ is the unit vector connecting two
neighboring spins $\mathbf{S}_i$ and $\mathbf{S}_j$. The general
phenomenological arguments of Ginzburg-Landau theory\cite{most06} as
well as symmetry arguments\cite{kaplan} showed that this coupling
between (ferro)electric polarization and magnetization always exists
in certain classes of materials independent of the crystal symmetry.
The MEE has been successfully described within the KNB-mechanism for
many classes of ferroelectric materials.
\cite{KNB_exp1,KNB_exp2,KNB_exp3,KNB_exp4} In turn, the discovery of
the deep relationship between magnetic structure and
ferroelectricity in several spin-chain materials stimulated further
research of multiferroics and paved a way between the physics of
multiferroics and quantum spin systems. The simplest system
exhibiting the MEE by means of KNB-mechanism, considered so far, is
the $S\!=\!1/2$ $J_1$-$J_2$-spin chain, which is believed to be a
more or less adequate model for several multiferroic materials such
as LiCu$_2$O$_2$,\cite{J1J2_1,J1J2_2,J1J2_3}
LiCuVO$_4$,\cite{J1J2_4,J1J2_5,J1J2_6} CuCl$_2$,\cite{J1J2_7} and
others. The expression for the electric polarization becomes
particularly simple in the case of a one-dimensional linear chain,
say, in $x$-direction. In this case, all $\mathbf{e}_{ij}=
\mathbf{e}_{i,i+1}= \mathbf{e}_x$, and one gets for the electric
polarization according to the KNB-mechanism:
\begin{subequations}\label{P_xyz}
\begin{align}
  P_x &= 0\epc\label{P_x}\\
  P_y &= \gamma\frac{1}{N}\sum_{n=1}^N
      \left(S_n^y S_{n+1}^x-S_n^x S_{n+1}^y\right)\epc\label{P_y}\\
  P_z &= \gamma\frac{1}{N}\sum_{n=1}^N
      \left(S_n^z S_{n+1}^x-S_n^x S_{n+1}^z\right)\epc\label{P_z}
\end{align}
\end{subequations}
where $\gamma$ is a material-dependent constant. Thus,
considering one of the simplest lattice spin models exhibiting some interplay
between magnetic and (ferro)electric properties, one arrives at the
Hamiltonian of a $J_1$-$J_2$-chain in a magnetic field with
additional Dzyaloshinskii-Moriya (DM)
interaction.\cite{KNB_exp1,KNB_exp2,KNB_exp3,KNB_exp4,J1J2_1,J1J2_2,J1J2_3,J1J2_4,J1J2_5,J1J2_6,J1J2_7}
Supposing the magnetic field to be in $z$-direction and the electric
field in $y$-direction, one gets:
\begin{align}\label{ham_J1_J2}
  \mcH &= J_1\sum_{n=1}^N \mathbf{S}_n \mathbf{S}_{n+1}+J_2
        \sum_{n=1}^N\mathbf{S}_n\mathbf{S}_{n+2}\\
  &\qd +\g E_y\sum_{n=1}^N\left(S_n^xS_{n+1}^y-S_n^yS_{n+1}^x\right)
        -g\mu_B H_z\sum_{n=1}^N S_n^z\epp \notag
\end{align}
Although the frustrated $S\!=\!1/2$ spin chain with ferromagnetic
nearest-neighbor and antiferromagnetic next-to-nearest-neighbor
interactions has recently received considerable amount of attention,
\cite{arl_2011,dmi_2011,kum_2012,kol_2012,ren_2012,sirker,jia11}
only few theoretical (mainly numerical) works have been devoted to
MEE-connected issues and to the interaction with an electric field
by means of DM-terms. The full Hamiltonian \eqref{ham_J1_J2} is not
integrable, whereas our aim is to develop an exact analysis of
the MEE. Therefore, we look at a slightly simpler model with only
nearest-neighbor interactions. Thus, we consider the $S\!=\!1/2$
XXZ-chain with DM-terms. We also suppose the potential ability of
the system to generate an electric polarization due to the
KNB-mechanism.
The integrability of the $S\!=\!1/2$ XXZ-chain with DM-terms is
based on a gauge transformation to the ordinary XXZ-chain. For the
latter we apply the method of the quantum transfer matrix (QTM)
leading to non-linear integral equations
(NLIE)\cite{qtm4,qtm7,qtm8,qtm9,qtm10,qtm11,qtm12} which determine
the exact free energy of the system, and thus, we are able to obtain
exact thermodynamic functions.

\section{The model}\label{sec2}

\subsection{Hamiltonian and MEE parameters}\label{sec2A}
Let us consider the $S\!=\!1/2$ XXZ-chain in a longitudinal
magnetic field and with Dzyaloshinskii-Moriya interaction:
\begin{align}\label{ham_XXZ_DM}
  \mcH &= J\sum_{n=1}^N\left(S_n^xS_{n+1}^x+S_n^yS_{n+1}^y
                    + \Delta S_n^zS_{n+1}^z \right)  \\
  &\qd + JE\sum_{n=1}^N\left(S_n^xS_{n+1}^y-S_n^yS_{n+1}^x\right)
                    -H\sum_{n=1}^NS_n^z\epp \notag
\end{align}
Here, periodic boundary conditions are assumed,
$S_{N+1}^{\alpha}=S_{1}^{\alpha}$, and the material-dependent
constants $\gamma$ and $g\mu_B$ are absorbed into the electric field
strength $JE=E_y\g$ and the magnetic field strength $H=g\mu_B H_z$,
respectively. The operators $S_n^{\alpha}$ obey the standard
$SU(2)$-algebra,
\begin{equation}\label{su2}
    \left[S_m^{\a},S_n^{\be}\right] = i \epsilon^{\a \be \g}S_m^{\g}\de_{mn}\epp
\end{equation}
Implementing the rotation of all spins about the $z$-axis
by angles proportional to the number of lattice sites,
\begin{equation}\label{GT1}
    S_n^{\pm} \rightarrow e^{\pm i n \phi}S_n^{\pm}\epc\qd \phi=\tan^{-1}(E)\epc
\end{equation}
one obtains the Hamiltonian of the ordinary XXZ-chain
with certain changes in $J$ and $\Delta$ and with twisted
boundary conditions. In other words, the gauge
transformation corresponding to the site-dependent
rotations of spins is equivalent to the following canonical
transformation of the initial Hamiltonian:
\cite{DM1,DM2, DM3,DM4}
\begin{align}
  \tilde{\mcH} &= \overline{\mcU}\mcH\mcU \label{UHU} \\
  &= \tilde{J}\sum_{n=1}^N\left(S_n^xS_{n+1}^x+S_n^yS_{n+1}^y
    +\tilde{\Delta}S_n^zS_{n+1}^z \right)-H\!\sum_{n=1}^N S_n^z\epc\notag\\
  \mcU &=\exp\left\{-i \phi\sum_{j=1}^N j S_j^z\right\}\epc\label{U}
\end{align}
where
\begin{equation}\label{JD(E)}
  \tilde{J} = J\sqrt{1+E^2}\epc \qd \tilde{\Delta} = \frac{\Delta}{\sqrt{1+E^2}}\epp
\end{equation}

As the Hamiltonian of Eq.~\eqref{UHU} is integrable and the methods
of QTM and NLIE provide a very efficient formalism to calculate
finite-temperature properties of the model, we are able to describe
the MEE by the influence of magnetic and electric fields on
thermodynamic functions. We are especially interested in the
magnetization and the electric polarization which are defined as
follows:
\begin{subequations}\label{MP}
\begin{align}
    M &= \frac{1}{N}\sum_{n=1}^N \big\langle S_n^z\big\rangle\epc \label{M}\\
    P &= \frac{1}{N}\sum_{n=1}^N \big\langle S_n^yS_{n+1}^x-S_n^xS_{n+1}^y\big
\rangle\epc \label{P}
\end{align}
\end{subequations}
Here, the angle brackets denote the thermal average with respect
to the statistical operator $\rho=\re^{-\mcH/T}/Z$. These
quantities can be derived from the canonical partition
function $Z=\text{tr}\big(\re^{-\mcH/T}\big)$ of the system.
Taking the connection between the Hamiltonians
\eqref{ham_XXZ_DM} and \eqref{UHU} into account, one can write down the
following relations:
\begin{subequations}\label{MP_XXZ}
\begin{align}
  M(T,H,E) &= -\left(\frac{\partial f}{\partial H}\right)_{T,E} = -\left(
\frac{\partial\tilde{f}}{\partial H}\right)_{T,\tilde{J},\tilde{\Delta}}\epc
\label{M_XXZ}\\[1ex]
  P(T,H,E) &= -\frac{1}{J}\left(\frac{\partial f}{\partial E}\right)_{T,H} =
-\frac{1}{J}\frac{d\tilde{J}}{d E} \left(\frac{\partial\tilde{f}}{\partial
\tilde{J}}\right)_{T,H,\tilde{J}\cdot\tilde{\Delta}}\notag\\ \label{P_XXZ}
  &= -\frac{E}{\sqrt{1+E^2}}\;G^{xy}(T,H,\tilde{J},\tilde{\Delta})\epc\\[1ex]
  \a(T,H,E) &= \frac{1}{J}\left(\frac{\partial M}{\partial E}\right)_{T,H} =
\left(\frac{\partial P}{\partial H}\right)_{T,E}\epc \label{a_XXZ}
\end{align}
\end{subequations}
where $\tilde{f}$ and $f$ are the free energies per lattice
site for the conventional XXZ-chain (Eq.~\eqref{UHU}) with
$E$-dependent coupling constant and anisotropy
(Eqs.~\eqref{JD(E)}) and for the XXZ-chain with
DM-interaction (Eq.~\eqref{ham_XXZ_DM}), respectively. Note
that the derivative in Eq.~\eqref{P_XXZ} with respect to
$\tilde J$ has to be performed at constant value of the
product $\tilde{J}\cdot\tilde{\Delta}$. Alternatively, we
may apply the canonical transformation \eqref{U} to the
definition \eqref{P} to obtain the result \eqref{P_XXZ}, where
$G^{xy}$ is the nearest-neighbor transverse
correlation function:
\begin{equation}\label{G}
    G^{xy}(T,H,\tilde{J},\tilde{\Delta}) = \frac{1}{N}
\sum_{n=1}^N\big\langle S_n^xS_{n+1}^x+S_n^yS_{n+1}^y\big\rangle\epp
\end{equation}
Here, angle brackets denote thermal averaging with
respect to the statistical operator
$\tilde{\r}=\re^{-\tilde{\mcH}/T}/\text{tr}\big(\re^{-\tilde{\mcH}/T}\big)$.

\subsection{Phase diagram}
\begin{figure}[!b]\begin{center}
\includegraphics[width=0.86\columnwidth]{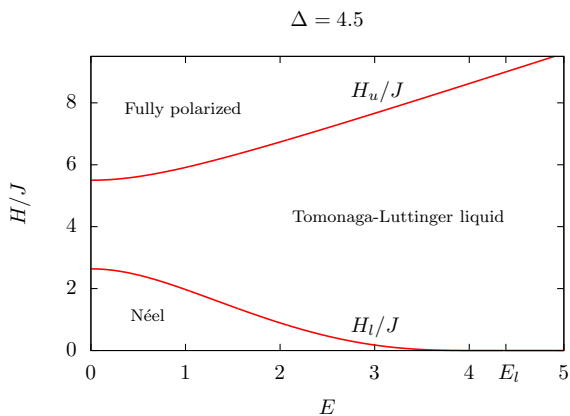}
\caption{\label{phase_diagram}(Color online)
Ground-state phase diagram of the $S\!=\!1/2$ XXZ-chain \eqref{ham_XXZ_DM}
for $\Delta=4.5$ in the $(E,H/J)$-plane. For $E=0$ the phase
transitions occur at $H_l/J=2.637\ldots$ and
$H_u/J=\Delta+1=5.5$. For $E>E_l$ there does not exist any
N\'eel ordering.}
\end{center}\end{figure}

The correspondence between $S\!=\!1/2$ XXZ-chains with and
without DM-interaction allows one to map the well known
ground-state phase diagram of the XXZ-chain onto the
$(E, H/J)$-plane (see Fig.~\ref{phase_diagram}). The
region between the two curves corresponds to the
massless polarized (antiferromagnetic) ground states, and
the region below the lower curve corresponds to the N\'eel
ordering of the spins. The following equations describe
the $E$-dependence of the upper and lower critical fields:
\begin{subequations}\label{H_c}
\begin{align}
  H_u/J &= \Delta + \sqrt{1+E^2}\epc \label{H_u}\\ \label{H_l}
  H_l/J &= \sinh{\h}\sum_{k=-\infty}^{\infty}\frac{(-1)^k}{\ch{(k\h)}}\epc
\end{align}
\end{subequations}
where $\ch{\h} = \tilde{\Delta} = \Delta/\sqrt{1+E^2}$ is
an appropriate parametrization of the anisotropy
$\tilde{\D}\geq 1$, i.\,e.~$E\leq E_l=\sqrt{\D^2-1}$. For
$E>E_l$ the electric field is too large for N\'eel ordering.

\section{Results}
\subsection{Free-fermion case}
As known, in case of zero anisotropy $\D=0$, the $S\!=\!1/2$
XXZ-chain reduces to the much simpler XX-chain which, in
turn, can be solved exactly by mapping it onto a free-fermion
system by means of the Jordan-Wigner transformation.\cite{JW}
The XX-chain with DM-term is also exactly solvable within
this context.\cite{der06a,der06b,der00,der99} Thus, starting
with the Hamiltonian \eqref{ham_XXZ_DM} with $\D=0$ and
performing a Jordan-Wigner transformation to free fermions:
\begin{subequations}\label{JWT}
\begin{align}
    S_n^+ &= \prod_{j=1}^{n-1}\left(1-2 c_j^+ c_j\right)c_n^+ \epc \label{Sn+}\\
    S_n^- &= \prod_{j=1}^{n-1}\left(1-2 c_j^+ c_j\right)c_n \epc \label{Sn-}\\
    S_n^z &= c_n^+c_n-1/2\epc \label{Snz}
\end{align}
\end{subequations}
where $c_n^+$ and $c_n$ are creation and annihilation
operators of spinless fermions obeying the standard
anti-commutation relations, $\left\{c_n,c_m\right\}=0$,
$\left\{c_n^+,c_m\right\}=\delta_{nm}$, one arrives at the
following free-fermion Hamiltonian:
\begin{align}\label{Ham_XX_DM.2}
  \mcH_{\text{XX-DM}} &= \frac{NH}{2} - H\sum_{n=1}^Nc_n^+c_n\\
  &\ +\frac{J}{2}\sum_{n=1}^N\left(\left(1+i E\right)c_n^+c_{n+1}+\left(1-i E\right)c_{n+1}^+c_n
\right)\epp \notag
\end{align}
Here, as usual, periodic boundary conditions are assumed.
Then, the Fourier transform diagonalizes the
Hamiltonian and yields
\begin{subequations}\label{Ham_XX_DM_FT}
\begin{align}\label{Ham_XX_DM_k}
\mcH_{\text{XX-DM}} &= \sum_{k}\varepsilon_k\left(c_k^+c_k-1/2\right)\epc \\
\varepsilon_k &= -H + J\left( \cos{k} + E\sin{k}\right) \label{eps_k}\epp
\end{align}
\end{subequations}

For the free-fermion picture of the model under
consideration, one can easily obtain all thermodynamic
functions in the thermodynamic limit in the form of
integrals over the one-dimensional Brillouin zone,
$k\in\left[-\pi,\pi\right]$ (see for example Ref.~[\onlinecite{der12}]).
So, one gets for the free energy $f$ per spin, the
magnetization $M$, the polarization $P$ and the
magnetoelectric tensor $\a$:
\begin{subequations}\label{therm_func}
\begin{align}
  f(T,H,E)  &= \frac{1}{2\pi}\int_{-\pi}^{\pi} \left(\frac{\varepsilon_k}{2}
                            +T\log(n_k)\right)\rd k\epc \label{free_energy_XX}\\
  M(T,H,E)  &= \frac{1}{2\pi}\int_{-\pi}^{\pi} (n_k-1/2)\,\rd k\epc \\
  P(T,H,E)  &= -\frac{1}{2\pi}\int_{-\pi}^{\pi} (n_k-1/2)\sin{k}\,\rd k\epc \\
  \a(T,H,E) &= -\frac{1}{2\pi T}\int_{-\pi}^{\pi} n_k(1-n_k)\sin{k}\,\rd k\epc
\end{align}
\end{subequations}
where $n_k=1/(1+e^{\varepsilon_k/T})$ are the occupation
numbers of spinless fermions.

\begin{figure}[!t]
\begin{center}
\includegraphics[width=0.86\columnwidth]{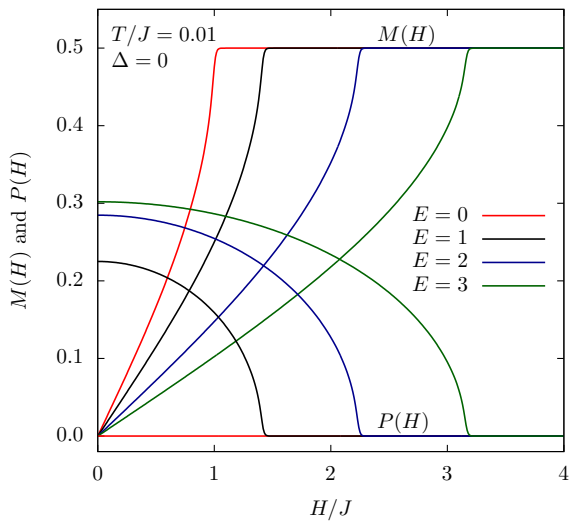}
\caption{\label{fcts_of_h_D=0_critical}(Color online)
Magnetization $M$ and polarization $P$ as functions of
the magnetic field $H$ for anisotropy $\Delta=0$,
temperature $T/J=0.01$ and different values of the electric
field $E=0,1,2,3$.}
\end{center}
\end{figure}

Figure \ref{fcts_of_h_D=0_critical} shows the dependence of
the magnetization $M$ and the polarization $P$ on the
magnetic field $H$ for different values of the electric
field $E$ in units of $J$ and for very low temperature. The
phase transitions at the corresponding values of
$H=H_u=\sqrt{1+E^2}$ can be observed. All critical
exponents are $1/2$. The curves in
Fig.~\ref{fcts_of_h_D=0_critical} are
obtained by the method described in the next section
and in App.~\ref{APP}, see for instance Eqs.~\eqref{MPexact}
and \eqref{shortrangecorrel}. Similarly, direct numerical
calculation of the same quantities by computing the
integrals on the r.\,h.\,s.~of Eqs.~\eqref{therm_func}
produce data with deviations smaller than $10^{-6}$.

\subsection{General case}
It is convenient to express the magnetization and the
polarization in terms of static short-range correlation
functions of the XXZ-chain. Using Eqs.~\eqref{M},
\eqref{P_XXZ}, and \eqref{G} as well as the translational
invariance of the Hamiltonian \eqref{UHU} one obtains
\begin{subequations}\label{MPexact}
\begin{align}
  M(T,H,E) &= \langle S_1^z \rangle\epc \label{Mexact}\\
  P(T,H,E) &= -\frac{E}{\sqrt{1+E^2}}\langle   S_1^xS_2^x+S_1^y S_2^y\rangle\epp \label{Pexact}
\end{align}
\end{subequations}
Due to recent progress in the theory of integrable models
static short-range correlation functions of the XXZ-chain
can be calculated exactly. It can be shown that all static
correlation functions of the one-dimensional XXZ-model can
be expressed as polynomials in the derivatives of three
functions\cite{JMS08} $\ph$, $\om$, and $\om'$ which are
determined by solutions of certain numerically well
behaved linear and non-linear integral equations.\cite{BoGo09}
In App.~\ref{APP} we give the definitions of these
functions in the massive case $\tilde{\Delta}>1$.\cite{TGK10a}
In the critical case $0\leq \tilde{\Delta}<1$ the
definitions are quite similar.\cite{BDGKSW08}

Employing the short-hand notations
\begin{subequations}
\begin{align}
  \ph_{(n)} &= \6_x^n \ph (x)|_{x=0}\epc\\
  f_{(m,n)} &= \6_x^m \6_y^n f(x, y)|_{x=y=0} \qd \text{for}\ f = \om, \om'\epc
\end{align}
\end{subequations}
we obtain\cite{TGK10a}
\begin{subequations}\label{shortrangecorrel}
\begin{align}
  \langle S_1^z\rangle &=-\2\ph_{(0)}\epc\label{s1zexact}\\ \label{s1xs2xexact}
  \langle S_1^x S_2^x\rangle &= \langle S_1^y S_2^y\rangle = \
        -\frac{\omega_{(0,0)}}{8 \sh{\h}}-\frac{\ch{\h}\ \omega'_{(1,0)}}{8\h}\epc
\end{align}
\end{subequations}
where the parameter $\h$ is defined by
$\tilde{\Delta}=\cosh{\eta}$ for
$\tilde{\Delta}>1$, i.\,e.~$E<\sqrt{\D^2-1}$. Similar
expressions can be obtained for the case
$0\leq\tilde{\Delta}<1$,\cite{BDGKSW08}
i.\,e.~$E>\sqrt{\D^2-1}$.

Combining Eqs.~\eqref{shortrangecorrel}, \eqref{MPexact},
and \eqref{a_XXZ} the magnetization, the polarization and
thereby the magnetoelectric tensor can be calculated
exactly for all temperatures over the whole range of the
phase diagram Fig.~\ref{phase_diagram}.

By solving the linear and non-linear integral equations of
Appendix \ref{APP}, the MEE parameters $M$, $P$ and $\a$ of
Eqs.~\eqref{Mexact}, \eqref{Pexact} and \eqref{a_XXZ},
respectively, can be calculated with high numerical accuracy.

\begin{figure}[!t]
\begin{center}
  \includegraphics[width=0.86\columnwidth]{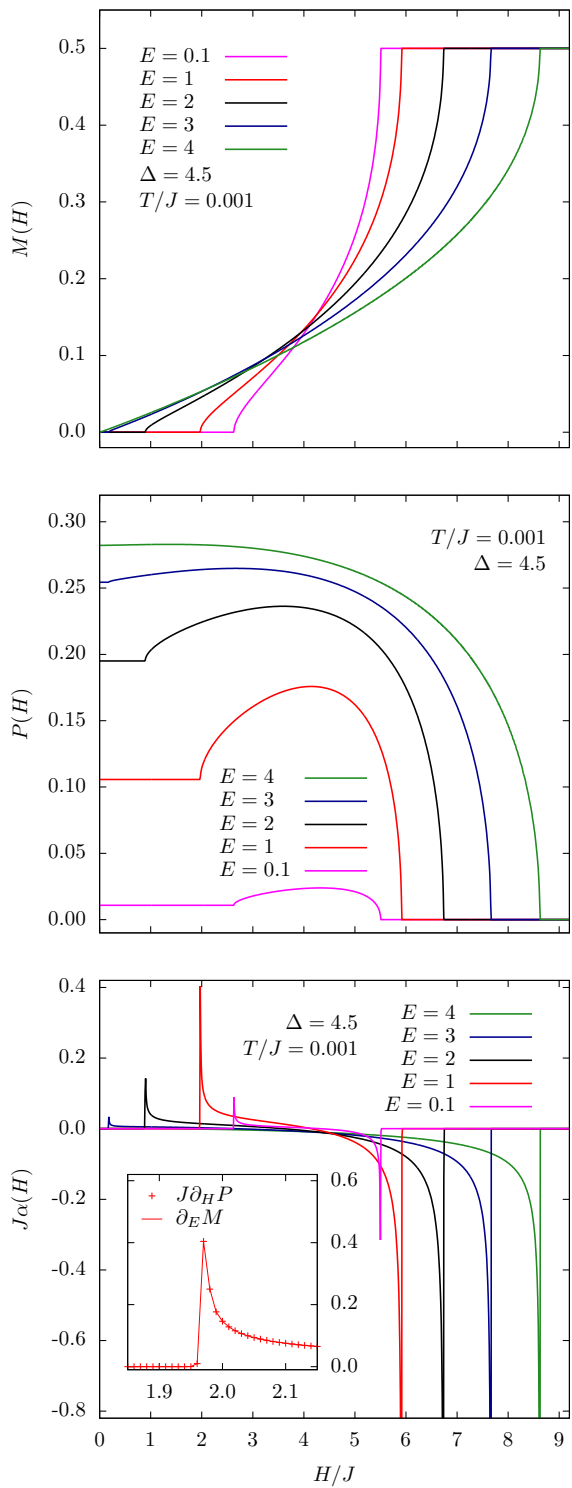}
\caption{\label{MPa_h_T1e-3}(Color online)
Magnetization $M$, polarization $P$, and magnetoelectric
tensor $\a$ as functions of the magnetic field $H$ for
anisotropy $\Delta=4.5$, temperature $T/J=0.001$ and
different values $E=0.1,1,2,3,4$. In the inset of the
third panel the two data sets $J\6_H P$ and $\6_E M$ are
compared, both obtained numerically using the five-point
stencil algorithm of the first derivative.}
\end{center}
\end{figure}

Figure \ref{MPa_h_T1e-3} shows these three quantities as functions
of the magnetic field for very low temperatures and different values
of the electric field. The phase transitions at the corresponding
values of the magnetic field (see Fig.~\ref{phase_diagram}) can be
easily observed. Interestingly, in the Tomonaga-Luttinger liquid
phase, the electric polarization $P$ is a non-monotonic function of
the magnetic field $H$ whereas it is constant, $P\neq 0$, in the
N\'eel-ordered phase and vanishing in the fully polarized
ferromagnetic phase. Here, the critical exponents characterizing the
magnetic field dependence of the electric polarization in the
vicinity of critical values of the magnetic field, $H=H_l(E)$ and
$H=H_u(E)$, are all equal to $1/2$,
\begin{subequations}
\begin{align}\label{eq:P_of_H_ind}
P &\sim\left(H-H_l\right)^{1/2}\epc \\
P &\sim\left(H_u-H\right)^{1/2}\epp
\end{align}\end{subequations}
This can be also observed by the van Hove-like singularities
of the magnetoelectric tensor $\a(H)$. We evaluated the
magnetoelectric tensor $\alpha(T,H,E)$ by numerical
derivatives of $P$ and $M$ with respect to $H$ and $E$,
respectively (see Eqs.~\eqref{alpha} or \eqref{a_XXZ}). In
the inset of Fig.~\ref{MPa_h_T1e-3} it is shown that both
data sets coincide.

\begin{figure}[!t]
\begin{center}
\includegraphics[width=0.86\columnwidth]{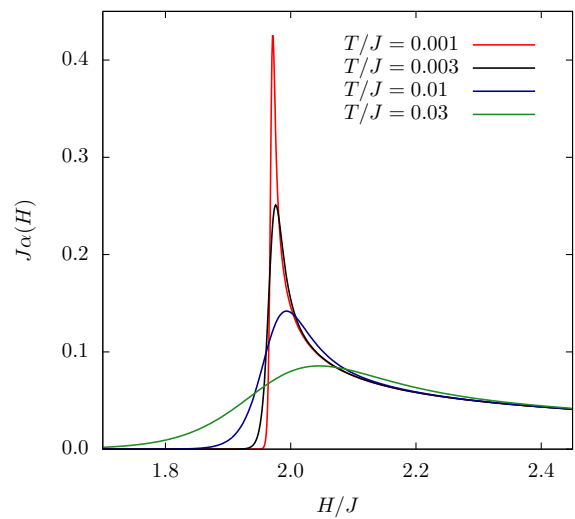} %
\caption{\label{alpha_of_h_E1_diffTs}(Color online)
 Temperature dependence of the magnetoelectric tensor $\a$
for $E=1$ and $\Delta=4.5$ close to the critical point $H\approx 1.95J$. The
square-root singularity for $T=0$ decays rapidly with
increasing temperature.}
\end{center}
\end{figure}

Figure \ref{alpha_of_h_E1_diffTs} illustrates the
temperature-driven melting of the singularity
of the magnetoelectric tensor as function of the magnetic
field. The values $E=1$ of the electric field and
$T=0.001\,J$ of the temperature belong to the red curve
of the third panel of Fig.~\ref{MPa_h_T1e-3}.
With increasing temperature the curve becomes more and
more flat. Note that the temperature in Fig.~\ref{alpha_of_h_E1_diffTs}
is only slightly increased. For $T=0.03\,J \ll J$, for
instance, there is no longer any indication of singular
behavior. The two curves $J\6_H P$ and $\6_E M$ would
overlay exactly such that only one of them is shown.

\begin{figure}[!t]
\begin{center}
\includegraphics[width=0.86\columnwidth]{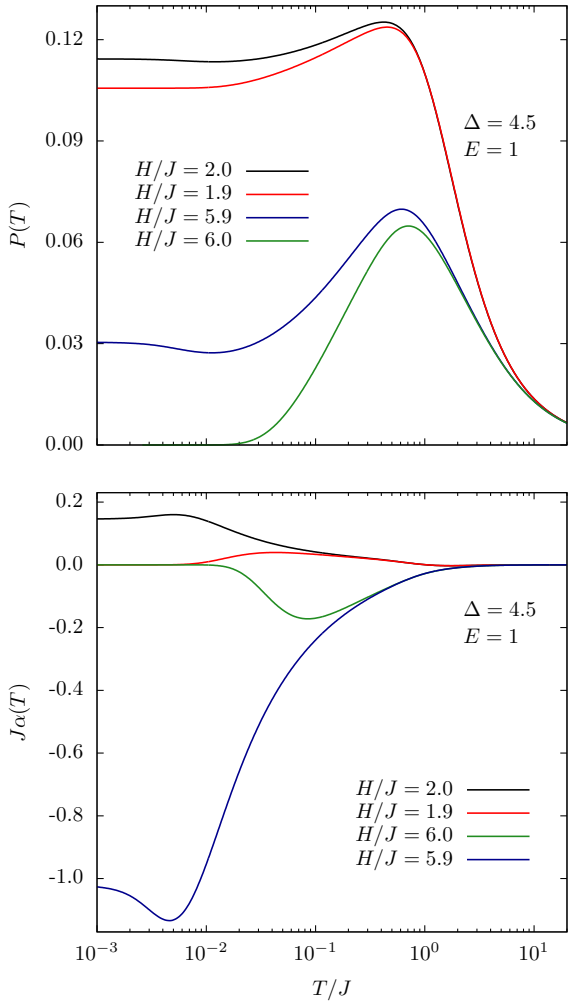}%
\caption{\label{alpha_of_T_E1_diffHs}(Color online)
Temperature dependence of the electric polarization $P$
and the magnetoelectric tensor $\a$
close to the critical points $H/J \approx 1.95$
and $H/J=4.5+\sqrt{2}$ (see Fig.~\ref{phase_diagram}).}
\end{center}
\end{figure}

In Figure \ref{alpha_of_T_E1_diffHs} the temperature
dependence of the electric polarization and the
magnetoelectric tensor are displayed.
The values of the electric and magnetic field are chosen
in a way such that the different ground-state phases are
represented (see Fig.~\ref{phase_diagram}), just below and
above the critical lines. All curves show non-monotonic
behavior and go to zero in the infinite-temperature limit.
In the zero-temperature limit the magnetoelectric tensor
vanishes in the N\'eel-ordered and in the fully polarized
ferromagnetic phase, whereas it remains constant in the
Tomonaga-Luttinger liquid.
\begin{figure}[!t]
\begin{center}
\includegraphics[width=0.86\columnwidth]{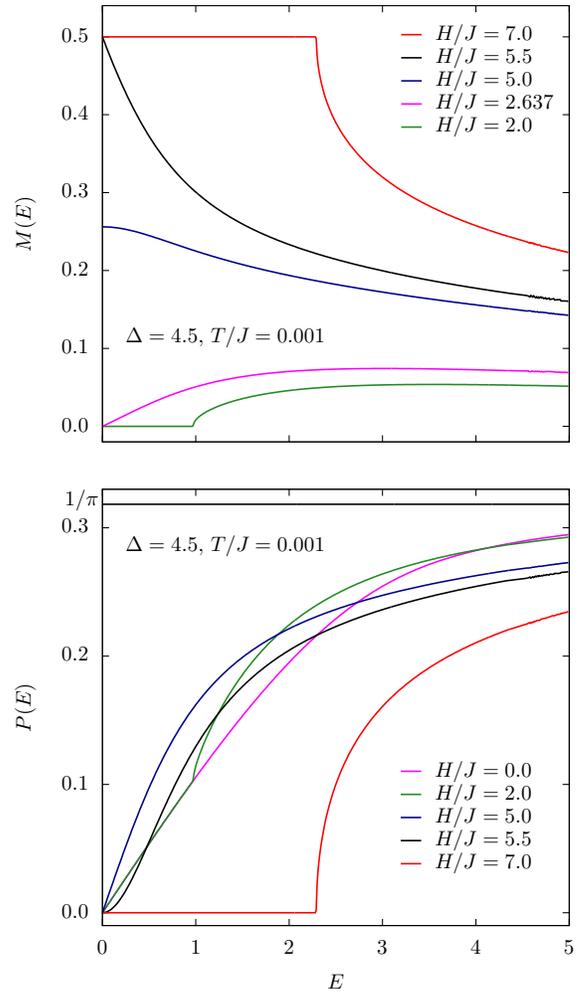} 
\caption{\label{MP_E_T1e-3_T1e-1}(Color online)
Electric field dependence of the magnetization $M$ and the
polarization $P$ for different fixed values of the magetic
field $H$ at very low temperature $T/J=0.001$. 
The values $H/J=2.637$ and $H/J=5.5$ correspond to the phase
transitions from the N\'eel-ordered ground state to the
Tomonaga-Luttinger liquid and from the Tomonaga-Luttinger
liquid to the fully polarized ferromagnetic state, respectively
(see Fig.~\ref{phase_diagram}).}
\end{center}
\end{figure}

Figure \ref{MP_E_T1e-3_T1e-1} shows the dependence of the
magnetization $M$ and the polarization $P$ on the electric
field $E$ for fixed values of the magnetic field $H$,
again for very low temperatures. For higher temperatures
the cusps smooth out. Phase transitions only
occur in the regimes (see Fig.~\ref{phase_diagram} with
$\D=4.5$)
\begin{subequations}
\begin{align}\label{eq:phasetrans}
    H/J < H_l(E=0)/J &= 2.637\ldots \epc \\
    H/J > H_u(E=0)/J &= \D+1 = 5.5 \epp
\end{align}
\end{subequations}
Here, the singularities are described by three different exponents.
In most cases one can observe square-root behavior. This is always
the case for the magnetization. However, for the special value of
magnetic field $H/J = 1+\D$, which corresponds to the transition
from the Tomonaga-Luttinger liquid phase to the fully polarized
ferromagnetic phase at zero electric field, the electric
polarization grows quadratically in $E$.

This can be explained as follows. The critical behavior of
the correlation function $\<S_1^x S_2^x\>$ for fixed $E$ as
function of the magnetic field $H$ is
$\<S_1^x S_2^x\> \sim \sqrt{H_u-H}$ for $H\leq H_u$. The gap
between the fixed value $H = \D+1$ and $H_u(E)= \D+\sqrt{1+E^2}$
is quadratic in $E$. Therefore, the critical behavior is
$\sqrt{H_u-H} \sim E$. Due to the prefactor in Eq.~\eqref{Pexact}
one eventually obtains
\begin{equation}
    P(E)\Big|_{H=\D+1} = -\frac{2E}{\sqrt{1+E^2}}\langle S_1^x S_2^x \rangle\Big|_{H=\D+1} \sim E^2\epp
\end{equation}
In addition to that, the polarization as a function of the external
electric field can also exhibit linear behavior with further
transition to the square-root form, at the values of $E$
corresponding to the transition from the N\'eel-ordered phase to
the Tomonaga-Luttiger liquid phase. Thus, one can define the
critical exponent $\delta_E$ which characterizes the behavior
of the polarization on its corresponding conjugate field
(electric field),
\begin{equation}
P\sim |E-E_c|^{\de_E}, \qqd \de_E=1/2,1,2\epp
\end{equation}

The limiting value of $P(E)$ for $E\to\infty$ can also be
understood. Due to Eqs.~\eqref{UHU} and \eqref{JD(E)} the
model reduces in this limit to a free-fermion model with
$H\to 0$. The rescaling $\tilde{J}\mapsto J$ yields
\begin{equation}\label{eq:HFF}
   \mcH \mapsto J\sum_{n=1}^N\left(S_n^x S_{n+1}^x + S_n^y S_{n+1}^y\right) \epp
\end{equation}
Then, the free energy per lattice site follows from
Eq.~\eqref{free_energy_XX} and is given by
\begin{equation}\label{therm_freeenergy}
  f(T,H,E)  = -\frac{T}{\pi}\int_{0}^{\pi} \log\left[2\,\text{cosh}{\left(\frac{J\cos{k}}{2T}\right)}\right]\,\rd k\epp
\end{equation}
Hence, the partition function $Z=\tr{(\re^{-H/T})}=\re^{-F/T}$
with $f=F/N$ and the nearest-neighbor correlation functions
can be calculated in the low-temperature limit,
\begin{align}
    \< S_1^x S_2^x \> =& -\frac{T}{2N}\frac{\6\log{(Z)}}{\6 J} = \frac{1}{2}\frac{\6  f}{\6 J}\notag\\
        =& -\frac{1}{4\pi}\int_{0}^{\pi}\tanh{\left(\frac{J\cos{k}}{2T}\right)}\cos{k}\,\rd k \notag\\
        \Rarrow[6mm]{\tiny $\!T\!\!\to\!0$}& -\frac{1}{4\pi}\int_{0}^{\pi}|\cos{k}|\,\rd k = -\frac{1}{2\pi}\epp
\end{align}
Therefore, $P(E) \to 1/\pi$ for $E\to\infty$ and $T\to 0$.

\section{Discussions and Conclusion}

The system considered in the present paper, being an integrable one, allowed us to obtain an exact description of the MEE by means of QTM and NLIE techniques. We considered a microscopic dielectric polarization to be proportional to the antisymmetric product of $x$- and $y$-components of the spins from two adjacent sites, i.\,e.~to a spin current $j^z$. This mechanism, which is known as a KNB-mechanism,\cite{KNB} is realized in several one-dimensional magnetic materials.\cite{KNB_exp1,KNB_exp2,KNB_exp3,KNB_exp4,J1J2_1,J1J2_2,J1J2_3,J1J2_4,J1J2_5,J1J2_6,J1J2_7} In our model, however, we have to impose a restriction in order to preserve integrability. While the standard model to describe the MEE in one-dimensional magnetic materials with KNB-mechanism is the $J_1$-$J_2$-chain, Eq.~\eqref{ham_J1_J2}, we consider an ordinary XXZ-chain in a magnetic field with DM-terms, Eq.~\eqref{ham_XXZ_DM}. The DM-terms describe a local dielectric polarization where the corresponding coefficient is the electric field magnitude $E$ in appropriate units.

The main objects of our investigation are the magneti\-zation $M$ and the di\-elec\-tric po\-lari\-zation $P$, which are the thermal averages of the operators $\sum_{n=1}^N S_n^z$ and $\sum_{n=1}^N(S_n^yS_{n+1}^x-S_n^xS_{n+1}^y)$, respectively. As a main result the exact finite-temperature plots of $M(T,H,E)$ and $P(T,H,E)$ have been obtained, especially in the low-temperature regime. Below we summarize the most interesting features of these plots.

As we mapped the system with DM-interaction onto the ordinary XXZ-chain, where $E$ enters the effective coupling constant $\tilde{J}(E)$ and effective anisotropy $\tilde{\Delta}(E)$, the magnetization curves $M(H)$ at fixed values of $T$ and $E$ (see Fig.~\ref{MPa_h_T1e-3}, upper panel) do not show any great difference from those of the conventional XXZ-chain.\cite{qtm4} The polarization as a function of magnetic field at constant values of temperature and electric field exhibits non-monotonic behavior (see Fig.~\ref{MPa_h_T1e-3}, middle panel).

Since we chose a large anisotropy $\Delta=4.5$, one can distinguish three possible parts of the polarization curve corresponding to the N\'eel-ordered phase, Tomonaga-Luttinger massless regime, and fully polarized spin configuration of the effective XXZ-chain, respectively. All these three regimes are easily recognizable in the middle panel of Fig.~\ref{MPa_h_T1e-3}. At low values of $H$, when the system is in the N\'eel-ordered ground state, the polarization  $P(H)$ is constant and thus starts with a plateau. One can also observe the effect of the electric field on the polarization $P(H)$. With increasing value of $E$ the plateau becomes shorter and higher. In the limit $E\rightarrow\infty$ the effective anisotropy parameter $\tilde{\Delta}=1/\sqrt{1+E^2}$ becomes zero, which means $M=0$ and $P=P_{max}=1/\pi$ for $H=0$. The second segment of the magnetic field dependent polarization corresponds to the Tomonaga-Luttinger regime of the effective XXZ-chain. This part of the polarization curve is a dome which starts at the plateau and ends at the value $P=0$ corresponding to the fully polarized ground state of the spin system. The values of the critical exponents corresponding to the two critical values of $H$ (transitions from the N\'eel-ordered ground state to the Tomonaga-Luttinger regime and from the Tomonaga-Luttinger regime to the fully polarized spin state) are equal to $1/2$.

A feature of the model under consideration is the vanishing polarization $P(H)$ at $E=0$. There only exists mutual influence of electric and magnetic properties of the model for $E\neq 0$. With respect to the electric properties the system does not show ferroic order, but is a conventional `dielectric' system which acquires polarization in an external electric field. However, this dielectric polarization is strongly affected by the external magnetic field as well, which is the MEE one can observe in our system.

Looking at the dependence of the magnetoelectric tensor $\a$ on the magnetic field $H$ (see Fig.~\ref{MPa_h_T1e-3}, lower panel) one can easily identify the zero-temperature phase transitions between N\'eel, Tomonaga-Luttinger and fully polarized ground states by the singular peaks. For sufficiently large $H$ the polarization is zero due to the transition to the fully polarized ferromagnetic state. The magnetoelectric tensor has a negative peak at this transition point. In addition to that, the absolute magnitude of this peak is larger than that of the positive peak corresponding to the transition from the N\'eel-ordered to the Tomonaga-Luttinger liquid state. This indicates a more dramatic drop of polarization for upper critical magnetic fields compared to its rise for the lower critical fields. However, the type of singularity in both cases is the same with critical exponent equal to 1/2, i.\,e.~$\a\sim \left|H - H_c \right|^{-1/2}$.

The behavior of the polarization as function of electric field is strongly affected by the magnetic field (see Fig.~\ref{MP_E_T1e-3_T1e-1}, lower panel). Even at vanishing magnetic field it exhibits non-trivial behavior during the whole process of polarization. The polarization increases linearly for small values of $E$ followed by a smooth curve approaching its saturated value $1/\pi$ in the limit $E \to \infty$. Switching on a magnetic field causes essential changes in the course of the polarization curve. For low magnetic fields the polarization starts linearly as in the $H=0$ case, but changes to a square-root behavior at the value of $E$ corresponding to the transition from N\'eel-ordered ground state to Tomonaga-Luttinger liquid phase. At this point the polarization curve has a cusp. For intermediate values of the magnetic field the system starts with a Tomonaga-Luttinger liquid phase, and the increase of $E$ does not drive the system to another ground state. One can observe monotonic increase of the polarization with linear behavior at small $E$. At the special value of $H$ corresponding to the critical value of the transition between Tomonaga-Luttinger liquid phase and fully polarized spin configuration at $E=0$, i.\,e.~$H=(\Delta+1)J$, one gets quadratic behavior for small $E$. Finally, for sufficiently large magnetic fields, when the system starts at the fully polarized magnetic configuration, the electric polarization vanishes. This changes dramatically to a square-root increase at the point of quantum phase transition to Tomonaga-Luttinger liquid state. To summarize one can distinguish four different behaviors of the polarization $P(E)$: linear followed by a cusp for $0< H < H_l$, linear with a smooth increase for $H=0$ or $H_l\leq H < H_u$, quadratic for $H=H_u$, and vanishing followed by a cusp for $H>H_l$.

Finally, the temperature dependencies of the polarization $P$ and of the magnetoelectric tensor $\a$ illustrate another interesting feature (see Fig.~\ref{alpha_of_T_E1_diffHs}). In the vicinity of quantum critical points the temperature dependence of the polarization is non-monotonic. The most intriguing feature shows the curve for $H=6J$ and $E=1$, where at low temperatures polarization is absent ($P=0$). Then, at intermediate temperatures the polarization curve rises up to values of about $0.065$. Eventually, it drops to zero for high temperatures. Thus, starting from zero temperature and increasing it continuously, the system exhibits a broad polarization peak over almost five orders of magnitude. Here, the values of magnetic and electric fields correspond to the fully polarized ferromagnetic phase just above the critical line. The temperature dependence of the magnetoelectric tensor, which is presented in the lower panel of Fig.~\ref{alpha_of_T_E1_diffHs}, is also non-monotonic and indicates the different regimes of the system response.

Although the system considered in the present paper exhibits MEE and also demonstrates a number of interesting features like many regimes of polarization and a thermally activated peak,
it does not exhibit ferroelectric polarization, which means that $P$ is always zero, unless $E\neq 0$. In order to obtain more interesting and richer features like double ferroic order or induction of polarization only by the magnetization and/or by the magnetic field within the KNB-mechanism, one should consider more sophisticated models. One way which seems straightforward to us is to include microscopic interactions between local order parameters, e.\,g.~local magnetization and local polarization. One example is
\begin{equation}\label{suz}
  \mcH_{int}\sim \sum_{n=1}^N \left( S_n^x S_{n+2}^y-S_n^y S_{n+2}^x \right)S_{n+1}^z\epp
\end{equation}
The local polarization is composed according to
KNB-mechanism and includes next-to-nearest-neighbor
spin interactions. Three-site interactions recently
received considerable attention in a bit diverse
context.\cite{der12,Suz71,Got99,Lou04,Kro08,Der09} It was shown by
Suzuki in Ref.~[\onlinecite{Suz71}] that there is a series of spin
chains with multiple-site interactions of special kind, which can be
mapped onto free spinless fermions via the Jordan-Wigner
transformation. Unfortunately, the $S\!=\!1/2$ XXZ-chain with three-site
spin-interaction terms given by Eq.~\eqref{suz} is no longer
integrable. One has to restrict oneself to the XY-chain with
corresponding terms\cite{der12,Suz71,Got99,Lou04,Kro08,Der09} or
has to implement numerical simulations. Another obvious non-trivial
choice of local interaction between magnetization and polarization
is $\sum_{n=1}^N p_n m_{n+1}$. However, in this case even the
corresponding XX-model is not exactly solvable.

Another possibility to construct the model exhibiting the double
ferroic order and MEE within the integrable systems is to consider
rather sophisticated spin chains with special three-spin interaction
discussed in Ref.~[\onlinecite{zvy03}]. These models seem to be very
promising by means of the MEE because of the local interaction
between microscopic magnetic and dipole moments presented in their
Hamiltonians (if one supposes that the KNB mechanism is realized in
the system). This argument allows us to hope that the model could
possess a ferroelectric phase generated by the magnetic field
solely. All these ideas conveyed above could be the guide line to
further developments in the MEE and multiferroics of integrable
systems.

\section{Acknowledgements}
The authors thank Jesko Sirker, Temo Vekua, Oleg Derzhko, Alexander
Nersesyan and Frank G\"ohmann for comments and stimulating
discussions. V.~O.~expresses his gratitude to the Department of
Theoretical Physics of the Bergische Universit\"at Wuppertal for
warm hospitality during the work on this paper. His research stay
has been supported by DFG (Grant No.~KL 645/8-1). V.~O.~also likes
to acknowledge partial financial support from Volks\-wagen
Foundation (Grant No.\ I/84 496) and from the project SCS-BFBR
11RB-001. M.~B.~likes to acknowledge financial support by
the DFG (Grant No.~KL 645/7-1).

\appendix

\section{Exact determination of correlation functions}\label{APP}
The functions $\ph$, $\om$, and $\om'$ that determine all
static correlation functions of the XXZ-chain are defined
in terms of solutions of linear and non-linear integral
equations. They were termed the physical part of the
problem,\cite{BoGo09} since the physical parameters like
temperature or magnetic field enter solely through these
functions. We provide their definitions only for the case
$\tilde{\Delta}>1$. The definitions for the case
$-1<\tilde{\Delta}<1$ can be found in Ref.~[\onlinecite{BDGKSW08}].

First of all let us define a basic pair of auxiliary
functions as the solution of the non-linear integral
equations\cite{TGK10a}
\begin{widetext}
\begin{subequations}\label{eq:nlie}
\begin{align}
  \log\mfb(x) &= -\frac{2\tilde{J}\sh(\h)}{T}d(x)-\frac{H}{2T}
+\kappa\ast\log\left(1+\mfb\right)(x)-\kappa^-\ast\log\left(1+\mfbq\right)(x)
\epc \label{eq:b_xxz}\\ \label{eq:bq_xxz}
  \log\mfbq(x) &= -\frac{2\tilde{J}\sh(\h)}{T}d(x)+\frac{H}{2T} +\kappa\ast
\log\left(1+\mfbq\right)(x)-\kappa^+\ast\log\left(1+\mfb\right)(x)\epc
\end{align}
\end{subequations}
\end{widetext}
where we denote with $\ast$ the convolution
$(f\ast g)(x)=\frac{1}{\pi}\int\nolimits_{-\pi/2}^{\pi/2}f(x-y)g(y)\rd y$.
Here we introduced the parameter $\h$ defined by $\tilde{\Delta} = \cosh(\h)$.
Eqs.~\eqref{eq:nlie} are valid for all $\h \ge 0$ meaning
that $\tilde{\Delta}>1$. The driving term $d$ is given by
its Fourier series,
\begin{equation}
  d(x) = \sum_{k=-\infty}^\infty \frac{\re^{2ikx}}{\ch(\eta k)} \epp
\end{equation}
Note that the physical parameters temperature $T$,
magnetic field $H$, and coupling $\tilde{J}$ enter only
through the driving terms of Eqs.~\eqref{eq:nlie} into
our formulae. The kernels $\kappa$ and $\kappa^\pm$ are
given by
\begin{subequations}
\begin{align}
    \kappa(x)& =\sum_{k=-\infty}^\infty\frac{\re^{-\h|k|+2i m kx}}{2\ch(\h k)}\epc\\
    \kappa^\pm(x)&=\kappa(x\pm i \eta^-)\epc
\end{align}
\end{subequations}
where $\eta^- =\eta-\eps$ with an arbitrary small number
$\eps>0$.

Except for the auxiliary functions $\fb$ and $\fbq$ we
need two more pairs of functions $g_\m^{(\pm)}$ and
${g'}_\m^{(\pm)}$ in order to define $\ph$, $\om$, and
$\om'$. Both pairs satisfy linear integral equations
involving $\fb$ and~$\fbq$,
\begin{widetext}
\begin{subequations}\label{eq:g_bb}
\begin{align}
  g^+_\nu(x) = -d(x-\nu)+\kappa\ast\frac{g^+_\nu}{1+\mfb^{-1}}(x)-\kappa^-
&\ast\frac{g^-_\nu}{1+\mfbq^{-1}}(x) \epc \\
  g^-_\nu(x) = -d(x-\nu)+\kappa\ast\frac{g^-_\nu}{1+\mfbq^{-1}}(x)-\kappa^+
&\ast\frac{g^+_\nu}{1+\mfb^{-1}}(x) \epc \\
  {g'}^+_\nu(x) = -c_+(x-\nu)+\kappa\ast\frac{{g'}^+_\nu}{1+\mfb^{-1}}(x)-\kappa^-
&\ast\frac{{g'}^-_\nu}{1+\mfbq^{-1}}(x)+\ l\ast\frac{g^+_\nu}{1+\mfb^{-1}}(x)-l^-
\ast\frac{g^-_\nu}{1+\mfbq^{-1}}(x)\epc \\
  {g'}^-_\nu(x) = -c_{-}(x-\nu)+\kappa\ast\frac{{g'}^-_\nu}{1+\mfbq^{-1}}(x)-\kappa^+
&\ast\frac{{g'}^+_\nu}{1+\mfb^{-1}}(x)+\
l\ast\frac{g^-_\nu}{1+\mfbq^{-1}}(x)-l^+
\ast\frac{g^+_\nu}{1+\mfb^{-1}}(x)\epp
\end{align}
\end{subequations}
The functions $l$ and $c_\pm$ are again given by their
Fourier series,
\begin{equation}
  l(x)=\sum_{k=-\infty}^\infty\frac{\sign(k)\re^{2ikx}}{4\ch^2(\h k)}\epc\qd\
  l^\pm(x)=l(x\pm i\eta^-)\epc\qd\
  c_\pm(x)=\pm\sum_{k=-\infty}^\infty\frac{\re^{\pm\h k+2ikx}}{2\ch^2(\h k)}\epc
\end{equation}
where we set $\sign(0)=0$ for the sign function.

The functions $\ph(\m)$, $\om(\m_1,\m_2)$, and
$\om'(\m_1,\m_2)$ that determine the explicit form of the
correlation functions of the XXZ-chain can be written as
integrals involving $\fb$, $\fbq$, $g_\m^{(\pm)}$, and
${g'}_\m^{(\pm)}$. The function
\begin{subequations}
\begin{align} \label{eq:phi_bb}
  \ph(\nu) &= \frac{1}{2\pi}\int\nolimits_{-\pi/2}^{\pi/2}\left(
\frac{g_{\nut}^-(x)}{1+\mfbq(x)^{-1}}-\frac{g_{\nut}^+(x)}{1+\mfb(x)^{-1}}
\right)\rd x\epc\\
\intertext{
determines the magnetization $M(T,H)= -\tst{\2}\ph(0)$
which is the only independent one-point function of the
XXZ-chain. The function}
\label{eq:omega_bb}
\omega(\nu_1,\nu_2) &= -4\kappa(\nut_2-\nut_1)+\tilde{K}_\eta(\nut_2-\nut_1)
- d\ast\left(\frac{g_{\nut_1}^+}{1+\mfb^{-1}}
+ \frac{g_{\nut_1}^-}{1+\mfbq^{-1}}\right)(\nut_2)\epc\\
\intertext{
also determines the energy per lattice site of the XXZ-chain,
$\< s_{j-1}^x s_j^x + s_{j-1}^ys_j^y + \D s_{j-1}^z s_j^z \> = \sh(\h) \om(0,0)/4$.
The function $\om' (\m_1, \m_2)$ is defined as}
\label{eq:omegaprime_bb}
  \frac{\omega'(\nu_1,\nu_2)}{\eta} &= -4 l(\nut_2-\nut_1)
+ \tilde{L}_\eta(\nut_2-\nut_1)-d\ast\left(\frac{{g'}_{\nut_1}^+}{1+\mfb^{-1}}
+ \frac{{g'}_{\nut_1}^-}{1+\mfbq^{-1}}\right)(\nut_2) \notag\\
& \mspace{200.mu}- c_{-}\ast\frac{g_{\nut_1}^+}{1+\mfb^{-1}}(\nut_2)
- c_{+}\ast\frac{g_{\nut_1}^-}{1+\mfbq^{-1}}(\nut_2)\epc
\end{align}
\end{subequations}
\end{widetext}
where we set $\nut=-i\nu$ and $\nut_j=-i\nu_j$. The
functions $\tilde{K}_\eta$ and $\tilde{L}_\eta$ are
determined by
\begin{subequations}
\begin{align}
    \label{eq:kschlange}
  \tilde{K}_\eta(x) &= \frac{\sh(2\eta)}{2\sin(x+i\eta)\sin(x-i\eta)}\epc \\
    \label{eq:lschlange}
  \tilde{L}_\eta(x) &= \frac{i\sin(2x)}{2\sin(x+i\eta)\sin(x-i\eta)} \epp
\end{align}
\end{subequations}
For the calculation of the magnetization and the
polarization, the non-linear integral equations for $\fb$
and $\fbq$ as well as their linear counterparts for $g_\m^{(\pm)}$
and ${g'}_\m^{(\pm)}$ were solved iteratively in Fourier
space utilizing the fast Fourier transformation algorithm.
The derivatives of $g_\m^{(\pm)}$ and ${g'}_\m^{(\pm)}$
with respect to $\m$, needed in the computation of the
derivative of $\om'$ satisfy linear integral equations
as well, which were obtained as derivatives of the
equations for $g_\m^{(\pm)}$ and ${g'}_\m^{(\pm)}$. Taking
into account derivatives is particularly simple in Fourier
space.

\end{document}